\documentclass[11pt,twoside]{article}
\usepackage{asp2010}

\resetcounters

\aspvoltitle{Galaxy Mergers in an Evolving Universe}
\aspcpryear{2012}

\begin{document}

\title{Driving the Gaseous Evolution of Massive Galaxies in the Early Universe}
\author{Dominik A.~Riechers$^1$
\affil{$^1$California Institute of Technology, 1200 East California Blvd, MC\,249-17, Pasadena, CA 91125, USA}}

\begin{abstract}
Studies of the molecular interstellar medium that fuels star formation and supermassive black hole growth in galaxies at cosmological distances have undergone tremendous progress over the past few years. Based on the detection of molecular gas in $>$120 galaxies at $z$=1 to 6.4, we have obtained detailed insight on how the amount and physical properties of this material in a galaxy are connected to its current star formation rate over a range of galaxy populations. Studies of the gas dynamics and morphology at high spatial resolution allow us to distinguish between gas-rich mergers in different stages along the ``merger sequence'' and disk galaxies. Observations of the most massive gas-rich starburst galaxies out to $z$$>$5 provide insight into the role of cosmic environment for the early growth of present-day massive spheroidal galaxies. Large-area submillimeter surveys have revealed a rare population of extremely far-infrared-luminous gas-rich high-redshift objects, which is dominated by strongly lensed, massive starburst galaxies. These discoveries have greatly improved our understanding of the role of molecular gas in the evolution of massive galaxies through cosmic time.
\end{abstract}

\section{Introduction}

Great progress has been made in studies of galaxy evolution out to
high redshift over the past years, but there are a number of
fundamental questions that remain to be answered. One of the most
important remaining issues is to understand whether star formation and
the subsequent buildup of stellar mass in galaxies at early cosmic
times occurs dominantly through major mergers (e.g., Springel et
al.\ 2005) or through a combination of minor mergers and steady,
so-called ``cold-mode'' accretion (e.g., Dekel et al.\ 2009). This
issue is closely connected to the question how galaxies obtain the
(dominantly molecular) gas that fuels star formation, and what their
gas mass fractions are. Gas mass fractions, in relation to star
formation rates, determine the evolutionary state of a galaxy (e.g.,
Daddi et al.\ 2010a; Tacconi et al.\ 2010). Next to these more general
issues, it is important to better understand the physical properties,
chemical composition, and dynamics of the star-forming gas in high-$z$
galaxies, which set the initial conditions for star formation (e.g.,
Riechers et al.\ 2006a, 2008a; Wei\ss\ et al.\ 2007).

All of these fundamental issues are directly tied to studies of
molecular gas in high redshift galaxies. Despite the great progress
that has been made in this field over the past 15\,years, our
understanding was ultimately limited by the technical capabilities of
past observatories in the centimeter to submillimeter wavelength
range, where the most common molecular gas diagnostics can be observed
at high redshift. New observatories, such as the Atacama Large
sub/Millimeter Array (ALMA) that currently nears completion, will be
key to ultimately solve many of the remaining mysteries in this
field.\ This article summarizes some of the most recent (pre-ALMA)
progress in the field of molecular gas observations out to high
redshift, and how the gas properties entangle with the driving
mechanism of star formation in massive high redshift galaxies.

\section{CO Detections at High Redshift:\ a Brief Summary}

%%%%%%%%%%%%%%%%%%%%%%%%%%%%%%%%%%%%%%%%%%%%%%%%%%%%%%%%%
%%%% Fig.1: Counts
%%%%%%%%%%%%%%%%%%%%%%%%%%%%%%%%%%%%%%%%%%%%%%%%%%%%%%%%%

\begin{figure}[h!]

\vspace*{-10.5mm}
      \begin{center}
%\hspace*{-60mm}
\includegraphics[width=12.5cm,angle=-0]{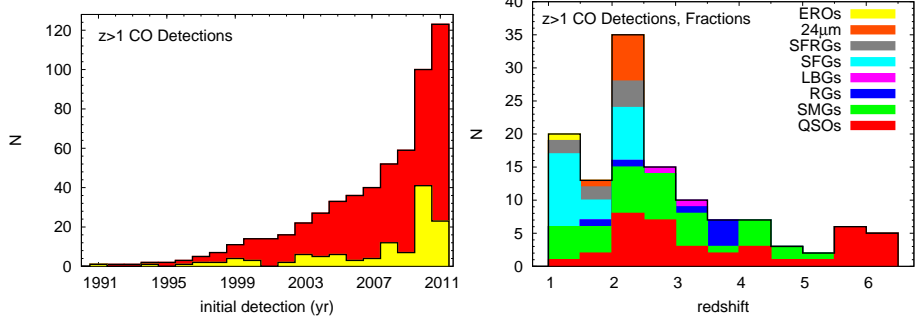}
      \end{center}
\vspace*{-8.5mm}

\caption{Detections of CO emission in $z$$>$1 galaxies. {\em
    Left}:\ Total number of detections (red) and detections per year
  (yellow) since the initial detection in 1991/1992. {\em
    Right}:\ Detections as of December 2011 as a function of redshift,
  and color encoded by galaxy type (figure updated from Riechers
  2011a).}
   \label{f1}
\vspace*{-3.5mm}

\end{figure}
%%%%%%%%%%%%%%%%%%%%%%%%%%%%%%%%%%%%%%%%%%%%%%%%%%%%%%

To date, molecular gas (most commonly CO) has been detected in $>$120
galaxies at $z$$>$1 (see reviews by Solomon \& Vanden Bout 2005; Omont
2007; Fig.~1), back to only 870\,million years after the Big Bang
(corresponding to $z$=6.42; e.g., Walter et al.\ 2003, 2004; Bertoldi
et al.\ 2003; Riechers et al.\ 2009). The bulk of these galaxies are
massive, hosting molecular gas reservoirs of $M_{\rm
  gas}$$>$10$^{10}$\,$M_\odot$, commonly with high gas fractions of
(at least) tens of per cent. The sensitivity of past observatories has
only allowed us to probe less massive and/or less gas-rich systems
with the aid of strong gravitational lensing (e.g., Baker et
al.\ 2004; Coppin et al.\ 2007; Riechers et al.\ 2010a, 2011a;
Swinbank et al.\ 2010, 2011).

Approximately 20\% of the detected systems are massive, gas-rich
optically/near-infrared selected star forming galaxies (SFGs; e.g.,
Daddi et al.\ 2010a; Tacconi et al.\ 2010), and 30\% each are
far-infrared-luminous, star-bursting quasars (QSOs; e.g., Wang et
al.\ 2010; Riechers 2011b) and submillimeter galaxies (SMGs; e.g.,
Greve et al.\ 2005; Tacconi et al.\ 2008; Fig.~1). The rest of
CO-detected high-redshift galaxies are limited samples of galaxies
selected through a variety of techniques, such as Extremely Red
Objects (EROs), Star-Forming Radio-selected Galaxies (SFRGs),
24\,$\mu$m-selected galaxies, gravitationally lensed Lyman-break
galaxies (LBGs), and radio galaxies (RGs; see Riechers 2011a for a
recent summary).

Besides CO, the high-density gas tracers HCN, HCO$^+$, HNC, CN, and
H$_2$O were detected towards a small subsample of these galaxies
(e.g., Solomon et al.\ 2003; Riechers et al.\ 2006b, 2007, 2010b,
2011b; Guelin et al.\ 2007; Omont et al.\ 2011).

\section{Luminosity Relations and the Star Formation Law}

The CO luminosity $L'_{\rm CO}$ is commonly considered to be a measure
for the total molecular gas mass $M_{\rm gas}$ in galaxies, and the
far-infrared (FIR) luminosity $L_{\rm FIR}$ is considered to be a
measure for the star formation rate (SFR; e.g., Solomon \& Vanden Bout
2005).  Thus, the relation between $L'_{\rm CO}$ or $M_{\rm gas}$ and
$L_{\rm FIR}$ or SFR may be considered a spatially integrated version
of the Schmidt-Kennicutt ``star formation law'' between gas surface
density and star formation rate surface density (e.g., Kennicutt
1998).

%%%%%%%%%%%%%%%%%%%%%%%%%%%%%%%%%%%%%%%%%%%%%%%%%%%%%%%%%
%%%% Fig.2: SF laws
%%%%%%%%%%%%%%%%%%%%%%%%%%%%%%%%%%%%%%%%%%%%%%%%%%%%%%%%%

\begin{figure}[h!]

\vspace*{-4mm}
      \begin{center}
%\hspace*{-60mm}
\includegraphics[width=11.5cm,angle=-0]{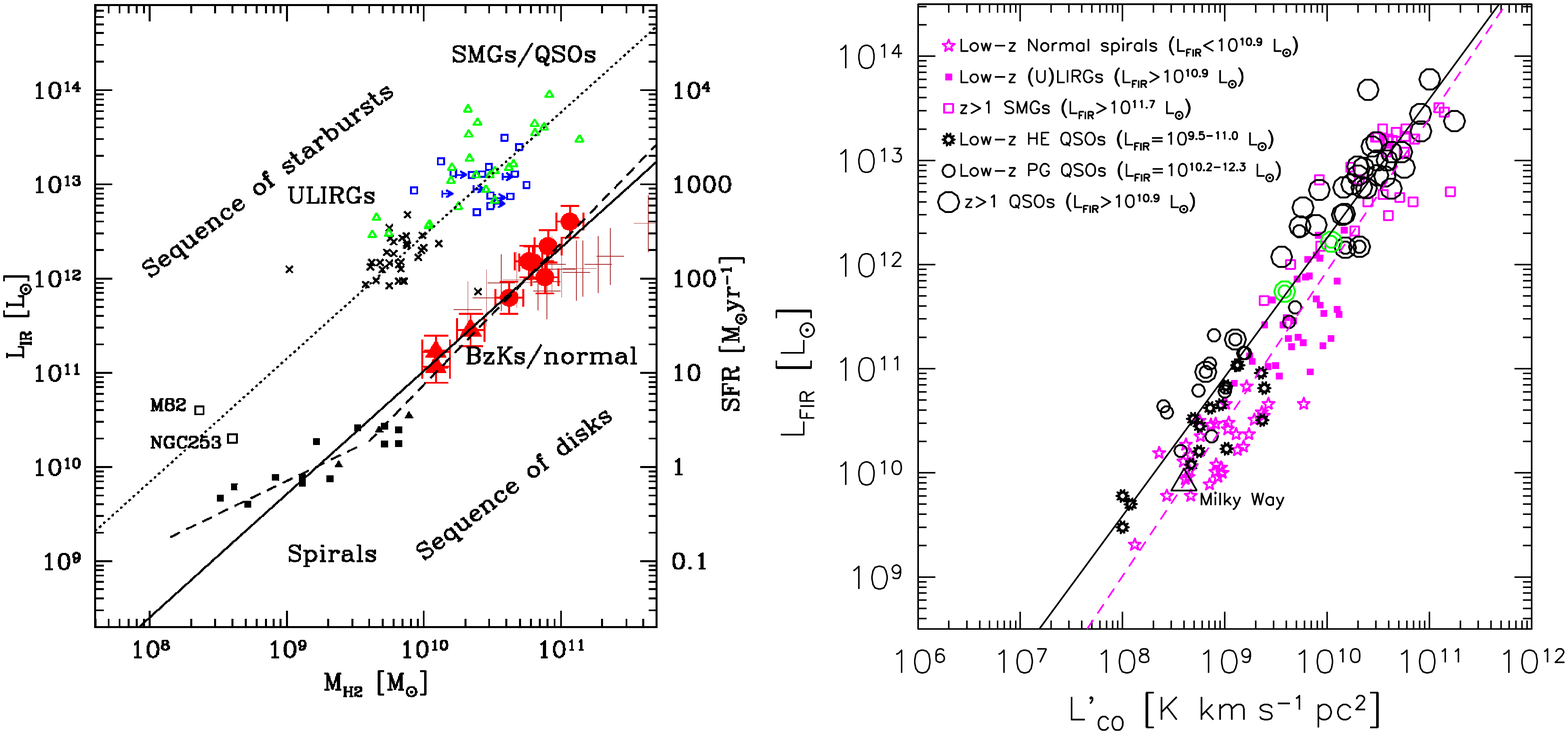}
      \end{center}
\vspace*{-8.5mm}

\caption{Comparison of CO luminosity/gas mass with (far-) infrared
  luminosity as a tracer of the star formation rate in low- and
  high-$z$ galaxies. {\em Left}:\ $M_{\rm gas}$--$L_{\rm IR}$ relation
  for low-$z$ spirals, starbursts and ULIRGs, and high-$z$ SFGs
  (``BzKs/normal''), SMGs, and quasars. The solid line represents a
  fit to spiral galaxies and SFGs, and the dotted line shows the same
  trend offset by 1.1dex in $L_{\rm IR}$ (which is consistent with the
  starburst galaxies, ULIRGs, SMGs and quasars). These two lines may
  represent two sequences for disk and starburst galaxies, with the
  offset being due to different dynamical timescales for star
  formation (e.g., Daddi et al.\ 2010b). {\em Right}:\ $L'_{\rm
    CO}$--$L_{\rm FIR}$ relation for quasars at low and high $z$
  (fitted by the solid line), and galaxies without dominant AGN at low
  $z$ (spirals and (U)LIRGs) and at high $z$ (SMGs; fitted by the
  dashed line). At high $L'_{\rm CO}$, galaxies with and without
  dominant AGN statistically occupy the same region, while at low
  $L'_{\rm CO}$, there is tentative evidence for an offset of quasars
  toward higher $L_{\rm FIR}$. This may indicate that AGN heating
  contributes significantly to $L_{\rm FIR}$ at low $L'_{\rm CO}$, but
  not at high $L'_{\rm CO}$ (Riechers 2011b).}
   \label{f4x}
\vspace*{-3.6mm}

\end{figure}
%%%%%%%%%%%%%%%%%%%%%%%%%%%%%%%%%%%%%%%%%%%%%%%%%%%%%%

Even in galaxies with luminous active galactic nuclei (AGN), $L_{\rm
  FIR}$ is commonly used as a proxy for the star formation rate in the
host galaxy. In principle, both the AGN and star formation can heat
the dust that gives rise to the continuum flux observed in the FIR,
but the characteristic dust temperatures of AGN heating are typically
by a factor of a few higher than those of dust heated by young
stars. Thus, the warm dust in AGN-starburst systems observed in the
rest-frame FIR is commonly thought to be dominated by heating within
the host galaxy, in particular in the most intense, dust-enshrouded
starbursts. If, however, $L_{\rm FIR}$ were to be dominated by the
AGN, one would expect an elevated $L_{\rm FIR}$ for such galaxies in
the $L'_{\rm CO}$--$L_{\rm FIR}$ relation. In Figure 2 (right panel),
a comparison of the $L'_{\rm CO}$--$L_{\rm FIR}$ relation for nearby
and high-$z$ quasars to nearby galaxies, (ultra-)luminous infrared
galaxies ((U)LIRGs) and SMGs without dominant AGN is shown (Riechers
2011b). For galaxies with low $L'_{\rm CO}$, there is an indication
for an excess in $L_{\rm FIR}$ for quasars relative to other systems;
however, there is no evidence for such a trend at high $L'_{\rm
  CO}$. This may suggest that, in systems with relatively low gas and
dust content, AGN contribute significantly to $L_{\rm FIR}$, but not
in systems with high gas and dust content. Thus, for massive
high-redshift galaxies, $L_{\rm FIR}$ appears to be a good proxy for
the SFR, even in quasars (Riechers 2011b).

Beyond the issue of dust heating, the question occurs if star
formation progresses the same way in all types of galaxies. In
disk-like spiral galaxies like the Milky Way, star formation takes
place in molecular clouds with compact, dense cores, confined by self
gravity (e.g., Solomon et al.\ 1987). In mergers, such as the Antennae
(NGC\,4038/39; Fig.~3), the gas and star formation appear to peak on
the galaxy nuclei, but also on relatively large scales in the dense
overlap region between the merging galaxies, leading to the formation
of so-called super star clusters (e.g., Wilson et al.\ 2003). Such
constellations are also commonly found in the highest-resolution CO
studies of high-redshift FIR-luminous massive galaxies (Fig.~3; e.g.,
Riechers et al.\ 2008a, 2008b).  In the nuclei of ULIRGs, the most
extreme nearby starbursts, star formation appears to occur in a dense,
intercloud medium, bound by the potential of the galaxy, rather than
in virialized clouds (e.g., Downes \& Solomon 1998). The differences
between starburst and disk galaxies are reflected in the star
formation law.  When comparing the $M_{\rm gas}$--$L_{\rm IR}$
relation for three largest CO-detected samples at high $z$, i.e.,
quasars, SMGs, and SFGs, to low redshift galaxies, two interesting
trends occur.  First, SFGs extend the relation found for nearby spiral
galaxies to higher $M_{\rm gas}$. Second, quasars and SMGs extend the
trend found for the most intense nearby starbursts and ULIRGs to
higher $M_{\rm gas}$. Both trends agree with the same slope, but are
offset by 1.1dex in $L_{\rm IR}$. Daddi et al.\ (2010b) interpret this
as evidence for two sequences in this relation for disk galaxies and
starbursts, which emerge from the different dynamical timescales of
star formation in these systems (Fig.~2, left; see also Genzel et
al.\ 2010). Recent theoretical studies have attempted to understand
the different trends based on the underlying conversion factor,
$\alpha_{\rm CO}$, from $L'_{\rm CO}$ to $M_{\rm gas}$, and find that
the two sequences could only be unified to a single relation when
assuming a broad continuum of conversion factors (e.g., Narayanan
2012, this volume).

%%%%%%%%%%%%%%%%%%%%%%%%%%%%%%%%%%%%%%%%%%%%%%%%%%%%%%%%%
%%%% Fig.3: Antennae vs. BRI1335
%%%%%%%%%%%%%%%%%%%%%%%%%%%%%%%%%%%%%%%%%%%%%%%%%%%%%%%%%

\begin{figure}[h!]

\vspace*{-6.5mm}
      \begin{center}
%\hspace*{-60mm}
\includegraphics[width=12.0cm,angle=-0]{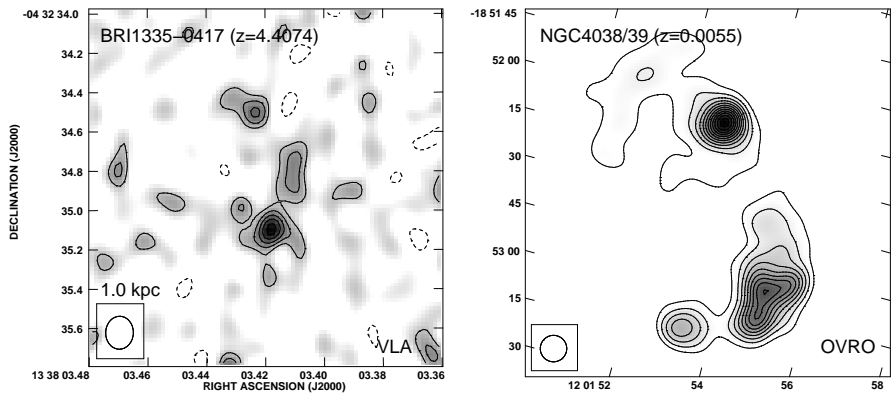}
      \end{center}
\vspace*{-15mm}

\caption{Molecular gas distribution in major mergers.
  {\em Left:}\ The $z$=4.41 AGN-starburst system BRI\,1335-0417
  (0.16$''$$\times$0.14$''$ resolution, corresponding to 1.0\,kpc;
  Riechers et al.\ 2008a). {\em Right:}\ The nearby major merger
  NGC\,4038/39 (the Antennae; convolved to 1.0\,kpc resolution and
  rotated for illustration; Wilson et al.\ 2003). Both systems are
  dominated by a clumpy, asymmetric gas distribution, and some diffuse
  emission is seen at low flux density level. It is plausible to
  assume that the clumps can be identified with two galaxy nuclei and
  a dense overlap region in both cases.}
   \label{f3}
\vspace*{-4.6mm}

\end{figure}
%%%%%%%%%%%%%%%%%%%%%%%%%%%%%%%%%%%%%%%%%%%%%%%%%%%%%%

\section{Gas Dynamics:\ A ``Merger Sequence'' of High Redshift Galaxies}

The advent of the Expanded Very Large Array (EVLA)\footnote{The
  Expanded Very Large Array was recently re-named to the Jansky VLA.}
has recently enabled studies of the full gas content, distribution and
dynamics of different populations of high-redshift galaxies through
observations of CO($J$=1$\to$0) emission (e.g., Riechers et
al.\ 2010a, 2011c, 2011d, 2011e; Ivison et al.\ 2011). In contrast to
previous studies at shorter wavelengths in higher rotational lines of
CO, observations of CO $J$=1$\to$0 trace the full amount and extent of
the molecular gas, and can be more directly compared to observations
in the nearby universe, which are most commonly undertaken in CO
$J$=1$\to$0.

One particular aspect of these studies is that they provide deeper
insight into the mechanisms that are driving the conversion of gas
into stellar mass in the most intensely star-forming galaxies at early
cosmic times, such as SMGs. It has been argued in the past that most
SMGs are major mergers of two gas-rich galaxies (e.g., Tacconi et
al.\ 2008; Engel et al.\ 2010).\ Detailed studies of CO($J$=1$\to$0)
emission with the EVLA confirm this picture, and show a range in
merger properties and stages (e.g., Riechers et al.\ 2011c, 2011d;
Fig.~4). Indeed, it becomes possible to place SMGs along a ``merger
sequence'' of high redshift galaxies, ranging from systems with two
disk-like gas-rich galaxies that are separated by tens of kiloparsec
in projection and several hundreds km\,s$^{-1}$ in velocity over
actively merging systems with a single, morphologically and
dynamically complex gas distribution to systems that have almost
reached coalescence. These early investigations demonstrate that
studies of larger samples at higher spatial resolution with the fully
upgraded EVLA in the future will allow us to distinguish between
different galaxy populations based on the physical properties of their
interstellar media. This will yield a more complete understanding of
the processes that trigger star formation and black hole activity in
the early universe, and their relative importance for the buildup of
stellar mass in galaxies as seen at present day.

%%%%%%%%%%%%%%%%%%%%%%%%%%%%%%%%%%%%%%%%%%%%%%%%%%%%%%%%%
%%%% Fig.4: SMG Merger Sequence
%%%%%%%%%%%%%%%%%%%%%%%%%%%%%%%%%%%%%%%%%%%%%%%%%%%%%%%%%

\begin{figure}[h!]

\vspace*{-9mm}
      \begin{center}
\hspace*{-3mm}
\includegraphics[width=12.0cm,angle=-0]{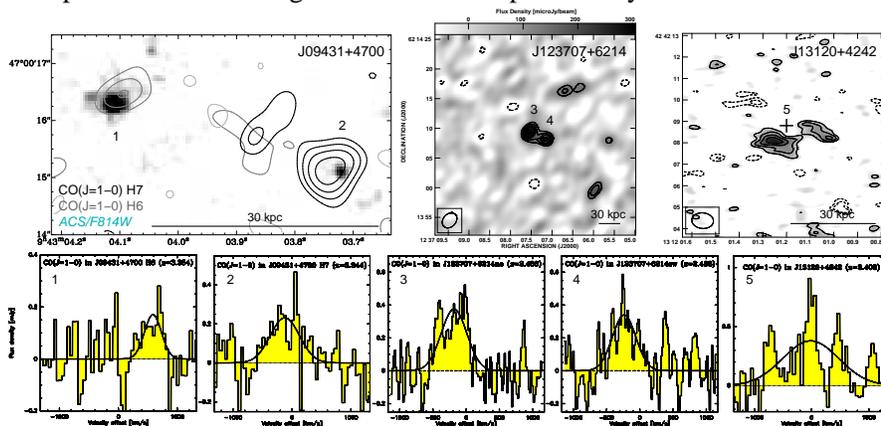}
      \end{center}
\vspace*{-8mm}

\caption{CO($J$=1$\to$0) observations of submillimeter galaxies along
  the ``merger sequence'' with the Expanded Very Large Array (Riechers
  et al.\ 2011c, 2011d). SMGs show complex gas morphologies that can
  extend over $>$10\,kpc scales, and commonly consist of multiple
  components. The gas distribution and kinematics of the majority of
  SMGs are as expected for major, gas-rich mergers. The observed
  diversity in these properties is consistent with different merger
  stages, as shown here for three examples. {\em Left}:\ The two
  gas-rich galaxies in this SMG are separated by tens of kpc and
  $\sim$700\,km\,s$^{-1}$, representing an early merger stage. {\em
    Middle}:\ This SMG still shows two separated components, but at
  similar velocity, representing a more advanced merger stage. {\em
    Right}:\ This SMG shows a single, complex extended gas structure
  with multiple velocity components, representing a fairly late merger
  stage.}
   \label{f4}
\vspace*{-7.6mm}

\end{figure}
%%%%%%%%%%%%%%%%%%%%%%%%%%%%%%%%%%%%%%%%%%%%%%%%%%%%%%

\section{Cosmic Environments of Massive Starbursts at Very High Redshift}

Recent studies have led to the discovery of the long sought-after
high-redshift tail of SMGs at $z$$>$4 (e.g., Capak et al.\ 2008; Daddi
et al.\ 2009). SMGs at the highest redshifts are of particular
interest, as they trace some of the most massive and active systems at
early cosmic times. In cosmological simulations, such systems are
expected to grow in the rare, most overdense regions in the early
universe (e.g., Springel et al.\ 2005). Thus, very high-$z$ SMGs may
trace the most distant proto-cluster regions in the universe, which
have the potential to grow into the most massive cosmic structures
seen at present day. Indeed, following the discovery of a SMG at an
unprecedented redshift of $z$=5.298, observational evidence has
recently been found for this mode of structure formation in the early
universe (Capak et al.\ 2011; Riechers et al.\ 2010c; Fig.~5). Deep
imaging and spectroscopy of its cosmic environment as part of the
Cosmic Evolution Survey (COSMOS) reveals a high overdensity of
Lyman-break galaxies at similar redshifts within a co-moving volume of
2\,Mpc radius around the SMG. The structure of the central
proto-cluster encompasses a halo mass of
$>$4$\times$10$^{11}$\,M$_\odot$, which would be consistent with what
is expected for the early formation of a present-day galaxy
cluster. Clearly, future discovery and studies of SMGs at comparable
and higher redshifts are desirable to determine the fraction of the
first massive starbursts that are associated with such galaxy
overdensities, as necessary to study the effects of environment on
galaxy formation within a more general cosmological context.

%%%%%%%%%%%%%%%%%%%%%%%%%%%%%%%%%%%%%%%%%%%%%%%%%%%%%%%%%
%%%% Fig.5: AzTEC-3
%%%%%%%%%%%%%%%%%%%%%%%%%%%%%%%%%%%%%%%%%%%%%%%%%%%%%%%%%

\begin{figure}[h!]

\vspace*{-3mm}
      \begin{center}
%\hspace*{-60mm}
\includegraphics[width=11.75cm,angle=-0]{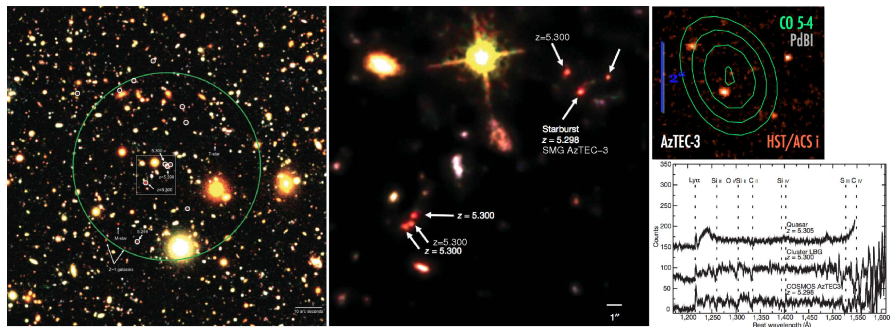}
      \end{center}
\vspace*{-8mm}

\caption{Gas properties and environment of the most distant SMG
  AzTEC-3 at $z$=5.298 (Riechers et al.\ 2010c; Capak et
  al.\ 2011). {\em Left}:\ 2$'$$\times$2$'$ region around AzTEC-3. The
  white circles show photometrically identified candidate Lyman-break
  galaxies (LBGs) at $z$=5.3, and the green circle indicates a region
  of 2\,Mpc co-moving radius. The labels indicate spectroscopic
  redshifts of confirmed LBGs or of low-$z$ galaxies and stars
  rejected based on their spectral energy distributions. {\em
    Middle}:\ Zoom-in of the central 0.865\,Mpc (co-moving) region,
  with spectroscopic redshifts labeled. {\em Right}:\ Zoom-in of the
  SMG, with CO($J$=5$\to$4) contours overlaid ({\em top}), and optical
  spectra of the SMG and companion galaxies ({\em bottom}). The
  massive, very high $z$ starburst galaxy AzTEC-3 traces a
  ``proto-cluster'' region of high galaxy overdensity in the early
  universe, showing that the highest-redshift SMGs may be an ideal
  tracer of the most massive cosmic structures at early cosmic times.}
   \label{f5}
\vspace*{-6.6mm}

\end{figure}
%%%%%%%%%%%%%%%%%%%%%%%%%%%%%%%%%%%%%%%%%%%%%%%%%%%%%%

\section{Herschel:\ The Lensing Revolution}

One of the most intriguing recent discoveries made by the {\em
  Herschel Space Observatory} is the identification of a rare
population of extremely luminous high-redshift starburst galaxies at
submillimeter wavelengths ($S_{500\mu m}$$>$100\,mJy; e.g., Negrello
et al.\ 2010). This population far exceeds the expected number counts
of SMGs at the bright end, and is dominated by gravitationally lensed
SMGs. This discovery is interesting, because it allows for a very
efficient selection of gravitational lenses based on a flux density
measurement alone (after rejection of low-redshift
contaminants). Also, the natural magnification in size and flux
density provided by gravitational lensing have enabled new techniques
to better understand the SMG population itself. First, the high
expected CO emission line fluxes, paired with recent spectral
bandwidth upgrades of radio/millimeter observatories, have yielded a
significant number of ``blind'' CO detections in these and similar
sources without any prior constraints on their redshifts (e.g.,
Riechers 2011b; Frayer et al.\ 2011; Cox et al.\ 2011; Harris et
al.\ 2012; Lupu et al.\ 2012). Second, high-resolution CO imaging
studies, in combination with detailed lens modeling, allow us to probe
down to structure sizes that would remain inaccessible with
current-generation instruments without the aid of gravitational
lensing (e.g., Fig.~6; Riechers et al.\ 2011a; Swinbank et al.\ 2011;
see also Riechers et al.\ 2008a).  Only ALMA will match the
sensitivity and resolution of these studies in unlensed galaxies, and
will even allow us to probe down to individual molecular cloud scales
in lensed systems.

%%%%%%%%%%%%%%%%%%%%%%%%%%%%%%%%%%%%%%%%%%%%%%%%%%%%%%%%%
%%%% Fig.6: Lock-01
%%%%%%%%%%%%%%%%%%%%%%%%%%%%%%%%%%%%%%%%%%%%%%%%%%%%%%%%%

\begin{figure}[h!]

\vspace*{-6mm}
      \begin{center}
%\hspace*{-60mm}
\includegraphics[width=11.5cm,angle=-0]{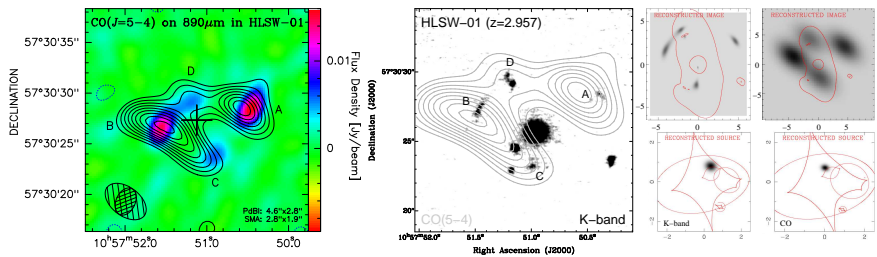}
      \end{center}
\vspace*{-8mm}

\caption{Multi-wavelength observations and modeling of HLSW-01, a
  rare, exceptionally bright, lensed SMG at $z$=2.957 discovered in
  the HerMES {\em Herschel}/SPIRE survey (Riechers et al.\ 2011a;
  Gavazzi et al.\ 2011). {\em Left}:\ Overlay of CO($J$=5$\to$4)
  contours on observed-frame 890\,$\mu$m continuum emission. A to D
  label the lensed images of the galaxy. {\em Middle}:\ Same overlaid
  on a 2.2\,$\mu$m image. {\em Right}:\ Lens model of the 2.2\,$\mu$m
  ({\em left}) and CO($J$=5$\to$4) emission ({\em right}) in the image
  ({\em top}) and source ({\em bottom}) planes. This SMG is lensed by
  a galaxy group at $z$$\simeq$0.6, yielding an extreme apparent
  250\,$\mu$m flux density of 425$\pm$10\,mJy due to a lensing
  magnification factor of $\sim$10.9, and a wide lens image separation
  of $\sim$9$''$. This source represents the bright end of a new
  population of lensed SMGs discovered with {\em Herschel}.}
   \label{f6}
\vspace*{-6.6mm}

\end{figure}
%%%%%%%%%%%%%%%%%%%%%%%%%%%%%%%%%%%%%%%%%%%%%%%%%%%%%%

\section{Summary and Outlook}

Studies of the molecular interstellar medium in galaxies at
cosmological distances have shown great progress in the past few
years. These advances were possible due to a combination of improved
selection techniques to identify gas-rich galaxies in the early
universe and major improvements in instrumentation, which have enabled
studies of less extreme galaxy populations than previously possible,
and have liberated high redshift molecular line studies from the
prerequisite of a precise spectroscopic redshift obtained through
other measures (which were a major selection bias in the past). The
most recent studies thus finally probe beyond the ``tip of the
iceberg'' of the most far-infrared-luminous galaxies with optical
spectroscopic redshifts that were the focus of molecular line studies
at high $z$ in the past. These observations also offer a ``sneak
peek'' into the detailed investigations of the physical properties of
galaxies at early cosmic times that will become possible with ALMA,
once it commences full science operations.

\acknowledgements I would like to thank the organizers of ``Galaxy
Mergers in an Evolving Universe'' for the invitation to this diverse
and stimulating conference. Also, I would like to thank my
collaborators on studies related to this subject, in particular Frank
Bertoldi, Peter Capak, Chris Carilli, Asantha Cooray, Pierre Cox,
Emanuele Daddi, Roberto Neri, Nick Scoville, Fabian Walter, and Ran
Wang. I acknowledge support from NASA through a {\em Spitzer Space
  Telescope} grant.\\[-6mm]

%\bibliography{aspauthor}

\end{document}